# Real-time Estimation of Vehicle Counts on Signalized Intersection Approaches Using Probe Vehicle Data

Mohammad A. Aljamal[1], *Member, IEEE,* Hossam M. Abdelghaffar[2], and Hesham A. Rakha[3], *Fellow, IEEE*

*Abstract*— **This paper presents a novel method for estimating the number of vehicles traveling along signalized approaches using probe vehicle data only. The proposed method uses the Kalman Filtering technique to produce reliable vehicle count estimates using real-time probe vehicle estimates of the expected travel times. The proposed method introduces a novel variable estimation interval that allows for higher estimation precision, as the updating time interval always contains a fixed number of probe vehicles. The proposed method is evaluated using empirical and simulated data, the former of which were collected along a signalized roadway in downtown Blacksburg, VA. Results indicate that vehicle count estimates produced by the proposed method are accurate. The paper also examines the model's accuracy when installing a single stationary sensor (e.g., loop detector), producing slight improvements especially when the probe vehicle market penetration rate is low. Finally, the paper investigates the sensitivity of the estimation model to traffic demand levels, showing that the model works better at higher demand levels given that more probe vehicles exist for the same market penetration rate.**

## I. INTRODUCTION

The number of on-road vehicles has increased rapidly over the past few decades. For example, in the United States, the number of motor vehicles registered from 1990 to 2016 increased by more than 75 million vehicles [1], leading to serious traffic congestion in many areas. One potential approach of mitigating traffic congestion is expanding the current infrastructure by adding new lanes and roadways to accommodate growing traffic demands; however, this comes with significant associated costs. A more efficient way of solving traffic congestion is improving traffic management strategies by using advanced technologies and algorithms.

Among the more recent technologies utilized for traffic management are Intelligent Transportation Systems (ITSs). ITS applications are developed to enhance transportation system efficiency, mobility, and reduce environmental impacts. In general, ITS aims at improving the infrastructure side of technology via sensors, communication, controllers, etc. Advanced Traffic Management Systems (ATMSs) constitute one ITS approach. ATMSs include advanced traffic signal control systems that optimize traffic signal timings in real-time [2-5].

Traffic density is defined as the number of vehicles on a given roadway segment divided by the length of the segment [6]. Knowing the number of vehicles on a specific roadway segment is crucial in developing efficient adaptive traffic signal controllers; however, it is difficult to measure traffic density directly in the field. Moreover, traffic occupancy measurements from loop detectors represent a temporal estimate of the traffic stream density around the measurement location. The research described herein attempts to estimate the number of vehicles, both queued and moving, along signalized roadway approaches (e.g., urban roadways) using only probe vehicle data. To the authors' knowledge, this work is the first attempt to estimate vehicle counts based solely on probe vehicle data. This estimation method will provide key input to real-time traffic signal controllers, leading to a reduction in intersection delays, vehicle emissions, and vehicle crashes.

## II. LITERATURE REVIEW

Past research has used different technologies/techniques, such as loop detectors [7-10], video detection systems [11], or data fusion techniques [12-15] (combining two different sources of data) to estimate the number of vehicles on signalized approaches. However, these techniques suffer from poor detection accuracy and have high installation costs. Emerging technologies, such as Global Positioning Systems (GPSs) and Connected Vehicle (CV) technologies, can provide and share vehicle real-time location and speed data.

[1] M. Aljamal is a Ph.D. Candidate in Civil Engineering. He is also with the Center for Sustainable Mobility at the Virginia Tech Transportation Institute (VTTI), Virginia Tech. (email: maljamal@vt.edu)

[2] H. Abdelghaffar is an Assistant Professor in the Department of Computers & Control Systems, Faculty of Engineering, Mansoura University, Egypt, and a Postdoctoral Associate in the Center for Sustainable Mobility at VTTI, Blacksburg, VA, 24061 USA.

[3] H. Rakha is the Samuel Reynolds Pritchard Professor of Engineering in the Charles E. Via, Jr. Department of Civil and Environmental Engineering and the Director of the Center for Sustainable Mobility at VTTI, Blacksburg, VA 24061 USA. (phone: 540-231-1505; fax: 540-231-1555; email: hrakha@vt.edu).

These sample data can be exploited and used to estimate the traffic density without the need to install additional hardware.

Numerous studies have attempted to estimate vehicle counts. For example, Ghosh and Knoop [7] demonstrated that vehicle counts can be improved by dividing the roadway into small segments (half-mile) to produce an efficient estimation. Another study used traffic flow and occupancy data from two conventional loop detectors to estimate the number of vehicles traveling along a specific road segment using the flow continuity equation [8]. Vigos *et al.* proposed a robust algorithm that requires at least three loop detectors on a roadway segment in order to estimate the number of vehicles; however, the cost of implementing such algorithms in the field is high. For example, imagine that the estimation is required for a city like New York, which has almost 12,460 signalized intersections [16]. If each intersection has four approaches, it would be necessary to install 150,000 detectors in order to make a proper estimate, which is an unreasonable proposition. Bhouri *et al.* proposed a scalar KF in order to estimate the number of vehicles on an on-ramp section, using a recorded film to observe the density measurements [10], utilizing the equation of conservation of cars as a state equation. Another study measured vehicle counts by matching vehicle signatures recorded by a network of wireless magnetic sensors [17]. An algorithm of video image processing was deployed by Beucher *et al.*, in which images were first collected from different scenes along a 150 m section, and then filters were used to detect the vehicle markers; however, this approach is difficult to utilize in the field.

Recently, data fusion has been widely used to estimate the number of vehicles along certain roadway sections, with the aim of achieving better accuracy than using only one source of data. In many of the works using data fusion, the Kalman Filtering (KF) technique [18] was employed for estimating traffic density. One study achieved accurate estimated traffic density results using the traffic flow values measured from a video detection system and the travel time obtained from vehicle GPSs [12]. The approach in this work is similar to that work in some aspects, but differs in two significant ways, namely: only probe vehicle data are used and furthermore the updating time interval is considered as a variable rather than a fixed value (the updating time interval was 1 minute in [12]). Van Erp *et al.* used data fusion to estimate the number of vehicles along an on-ramp segment [14]. They used traffic flow data from loop detectors and aggregated speeds from floating cars, the latter provided by Google, and set a 300s fixed updating interval. Another study used loop detectors and probe vehicle data to estimate freeway traffic density [15], relying on IntelliDrive technology (vehicle infrastructure integration) at a predefined updating time interval. Anand *et al.* used video and GPS data to estimate the number of vehicles along a roadway segment [13]. In that study, video captured the traffic flow at the segment's entrance and exit points, while GPS data provided travel time measurements.

Several researchers have used the KF technique to enhance vehicle count estimates in various transportation applications, such as speed [19, 20], travel time [21, 22], and traffic flow [23]. An unscented KF deployed for speed estimation using single loop detectors [19] with a nonlinear state-space equation, was able to improve the speed estimates. Another study employed a linear KF technique to estimate speed, relying on the relationship between the flow-occupancy ratio and vehicle speed [20], yielding acceptable speed estimates for congested traffic conditions. A cumulative travel-time responsive (CTR) real-time intersection control within a CV environment was also developed using the KF technique [21]. In that study, the authors recommended having at least 30% levels of market penetration (LMPs) in order to realize the CTR algorithm's benefits.

In summary, the existing literature showed the benefits of using the KF technique to reduce errors and address different aspects of the traffic state estimation problem. Accordingly, the KF technique was adopted in this study. One commonality of the aforementioned studies is that they all estimated the number of vehicles using one source of data from fixed sensors (e.g., loop detectors) or using fused source data (e.g., video with GPS data) utilizing a predefined updating interval.

In this research, a scalar KF technique was applied to estimate real-time vehicle counts along signalized approaches using both real and simulated traffic data. This estimation technique was applied to a signalized approach in downtown Blacksburg, VA. The proposed algorithm extends the-state-of-the-art in vehicle count estimates by making four major contributions:

1. Past research used a fixed estimation time interval (e.g., 20 s). However, using a fixed estimation period might lead to inaccurate estimation, especially at low LMPs; however, treating the estimation interval as a variable leads to improved estimation. In this work, we defined the estimation time interval as the point at which exactly *n* probe vehicles traversed the tested approach (see Observer 2 in Fig. 3).
2. This study relies only on probe vehicle data. Different probe vehicle LMPs were also tested (ranging from 10% to 90% in increments of 10%), and recommendations for future research are presented.
3. This study examines the estimation accuracy when adding a single loop detector, and a sensitivity analysis is made in terms of the optimum location of the stationary sensor.
4. This study investigates the sensitivity of the proposed estimation model to different factors including the length of the approach, the level of traffic congestion, and the LMP of probe vehicles.

This paper is organized into four additional sections. The first section describes the estimation method and the problem formulation. The second section describes the data collected from downtown Blacksburg, VA. The third section shows the results of the new proposed model. The fourth section presents the conclusions of the study and recommended future work.

## III. METHODS

### A. Define the Estimation Interval Time

Unlike other studies, this work defined the estimation time interval as a variable rather than a fixed value. This new approach enhances the estimation at low LMPs, as shown later in the Results section. For example, if the approach's LMP is 10%, the number of probe vehicles will obviously be low. If we treat the problem using a fixed estimation interval, then the probability of observing zero probe vehicles within an interval will be high for short estimation interval durations, making the estimation inefficient and inaccurate. Accordingly, low LMPs require long intervals (e.g., 300 s) to ensure that at least one probe vehicle is on the approach. In contrast, approaches with high LMPs can use short estimation intervals (e.g., 20 s). One of the major contributions of this study was to address this issue to produce an efficient and convenient way of determining the duration of the estimation period.

For this work, the updating time interval was defined as the time when an exact number of probe vehicles traversed the approach (i.e. reached the traffic signal stop bar)—reflecting a pre-defined sample size ($n$). This new approach ensures that the same number of probe vehicles are used for each updating time interval. Thus, using this approach, the population confidence interval will be a fixed scaling of the sample confidence interval. This approach also ensures that there will be sufficient information about the probe vehicles. Consequently, the average travel time ($TT$) value in Equation (13) will always be observed.

### B. Formulation

This section defines the proposed formulation to estimate the total number of vehicles along a signalized approach. The proposed method utilizes the KF technique, which is comprised of two equations: (a) a state equation and (b) a measurement equation. The state equation is based on the traffic flow continuity equation as defined in Equation (1), while the measurement equation is based on the hydrodynamic relation of traffic flow given in Equation (3). The KF is a recursive estimation model that continuously repeats the state estimations and corrections.

Equation (1) computes the number of vehicles by continuously adding the difference in the number of vehicles entering and exiting the section to the previously computed cumulative number of vehicles traveling along the section. This integral results in an accumulation of error that requires fixing, and thus the need for the measurement equation.

$$N(t) = N(t - \Delta t) + \frac{\Delta t}{\rho}[q^{in}(t) - q^{out}(t)] \quad (1)$$

Here $N(t)$ is the number of vehicles traversing the approach at time ($t$), $\Delta t$ is the duration of the variable updating time interval, $N(t-\Delta t)$ is the number of vehicles traversing the approach in the previous interval, $q^{in}$ and $q^{out}$ are the probe flows entering and exiting the approach between ($t-\Delta t$) and ($t$), respectively, and $\rho$ is the LMP of probe vehicles. The equation above produces accurate results if the scaled traffic flows ($q^{in}/\rho^{in}$ and $q^{out}/\rho^{out}$) are accurate [24, 25]. The total counts can be extracted from traditional loop detectors or video detection systems. We should note here that the $\rho$ value in Equation (1) plays a major role in delivering accurate outcomes. The $\rho$ is defined as the ratio of the number of probe vehicles ($N_{probe}$) to the total number of vehicles ($N_{total}$), as shown in Equation (2). For instance, if $\rho$ is 0.5, and the number of probe vehicles is 5, then the expected total number of vehicles is 10.

$$\rho = \frac{N_{probe}}{N_{total}} \quad (2)$$

Equation (3) describes the hydrodynamic relationship between the macroscopic traffic stream parameters (flow, density, and space-mean speed).

$$q = ku \quad (3)$$

Where $q$ is the traffic flow (vehicles per unit time), $k$ is the traffic stream density (vehicles per unit distance), and $u$ is the space-mean speed (distance per unit time). The space-mean speed can be replaced using Equation (4),

$$u = D/TT \quad (4)$$

where $D$ is the approach length, and $TT$ is the average vehicle travel time. Since probe vehicles can share their instantaneous locations every $\Delta t$, the travel time of each probe vehicle can be computed for any road section. Thus, the probe vehicle travel time is used in the measurement equation, as shown in Equations (5), (6), and (7).

$$TT(t) = D \times \frac{k(t)}{\bar{q}(t)} \quad (5)$$

$$TT(t) = \frac{1}{\bar{q}(t)}[k(t) \times D] = \frac{1}{\bar{q}(t)} N(t) \quad (6)$$

$$TT(t) = H(t) \times N(t) \quad (7)$$

Where $\bar{q}$ is the average traffic flow entering and exiting the approach, and $H(t)$ is a transition vector that converts the vehicle counts to the average travel time. $H(t)$ is the inverse of the average flow (i.e. the first term of Equation (6)), as shown in Equation (8).

$$H(t) = \frac{1}{\bar{q}(t)} = \frac{2 \times \rho}{q^{in}(t) + q^{out}(t)} \quad (8)$$

The proposed estimation method can be solved using the KF equations, as follows:

$$\widehat{N}^-(t) = \widehat{N}^+(t - \Delta t) + \frac{\Delta t}{\rho}[q^{in}(t) - q^{out}(t)] \quad (9)$$

$$\widehat{TT}(t) = H(t) \times \widehat{N}^-(t) \quad (10)$$

$$\widehat{P}^-(t) = \widehat{P}^+(t - \Delta t) \quad (11)$$

$$G(t) = \frac{\widehat{P}^-(t) H(t)^T}{H(t) \widehat{P}^-(t) H(t)^T + R} \quad (12)$$

$$\widehat{N}^+(t) = \widehat{N}^-(t) + G(t) [TT(t) - \widehat{TT}(t)] \quad (13)$$

$$\hat{P}^+(t) = \hat{P}^-(t) \times [1 - H(t)G(t)] \qquad (14)$$

Where $\hat{N}^-$ is the a priori estimate of the vehicle counts calculated using the measurement prior to instant $t$, and $\hat{P}^-$ is the a priori estimate of the covariance error at instant $t$. The Kalman gain ($G$) is computed as the ratio of the state error to the sum of the state errors plus the measurement, as demonstrated in Equation (12). The $R$ variable is the covariance error of the measurements. The posterior state estimate $\hat{N}^+$ and the posterior error covariance estimate $\hat{P}^+$ are updated using Equations (13) and (14) after considering the probe vehicle travel time measurements.

It should be noted here that the $\rho$ value in Equation (9) impacts the estimation accuracy significantly especially for low LMPs (e.g., LMP<30%) given that it scales the estimated values. For instance, for a $\rho$ value of 0.1, the second term of Equation (9) is scaled by a factor of 10, which results in large errors in the vehicle counts. A real illustrative example using empirical data is described in TABLE I. This example shows the impact of low LMPs in the state equation (i.e., $\rho = 0.1$). For the first estimation step, the total number of arrivals ($A_T$) and departures ($D_T$) within the polling interval are 65 and 60 as displayed in TABLE I, whereas the number of probe vehicle arrivals ($A_P$) and departures ($D_P$) for the same polling interval are 6 and 5, respectively. The first estimation starts with an erroneous initial vehicle count estimate $\hat{N}^+(0) = 5$ vehicles while the real number is zero as is done in [9], the actual total number of vehicles on the approach would then be 5 vehicles (0+(65-60)). Applying Equation (9) would produce an estimated total of 15 (5+(6-5)/0.1=15). Note that if the $\rho$ value in Equation (9) is constrained by a lower bound it can prevent the state equation from producing such large errors. For the same example, if the $\rho$ lower bound in Equation (9) is set to be equal 0.5, the total number of vehicles is estimated to be 7 (5+(6-5)/0.5). As can be seen from this example, the absolute error using a lower bound of 0.5 is 2 vehicles, whereas the absolute error using the estimated total counts is 10 vehicles. Consequently, it is much easier for the KF to correct a small error rather than a large error after applying the measurement equation. Fig. 1 compares the vehicle counts using the state equation; Equation (9) with and without a lower bound on $\rho$, along the estimation steps for the 10% LMP scenario. It is clear that having a lower bound on $\rho$ in the state equation improves its accuracy by reducing the distance from the actual vehicle count line (the green line in Fig. 1). The reason behind not allowing the market penetration to be very small is that: (1) we use a single $\rho$ value estimate to approximate the two $\rho$ values (upstream and downstream of the approach); and (2) if the $\rho$ value is very small, the approximation error is inflated, producing a larger error in the estimated total number of vehicles. Consequently, Equation (9) can be re-written as presented in Equation (15). The $\rho$ factor in our state equation becomes the maximum of the historic $\rho$ value and a lower bound ($\rho_{min}$), taken to be 0.5 in our analysis. Note that further work in the future is needed to develop procedures to provide more accurate estimates of the two market penetration rates; namely, the one at the entrance of the approach and the one at the exit of the approach.

$$\hat{N}^-(t) = \hat{N}^+(t - \Delta t) + \Delta t \frac{q^{in}(t) - q^{out}(t)}{\max(\rho, \rho_{min})} \qquad (15)$$

TABLE I
VEHICLE COUNT ESTIMATION VALUES OF USING EQUATIONS (9) AND (15)

| Estimation Step | Probe Arrival ($A_P$) | Probe Departure ($D_P$) | Total Arrival ($A_T$) | Total Departure ($D_T$) | $\hat{N}^-$ (Eq. 9) | $\hat{N}^-$ (Eq. 15) | Actual $N_T(t - \Delta t) + A_T - D_T$ |
|---|---|---|---|---|---|---|---|
| 1 | 6 | 5 | 65 | 60 | 15 | 7 | 5 |
| 2 | 4 | 5 | 67 | 69 | 5 | 5 | 3 |
| 3 | 6 | 5 | 66 | 62 | 15 | 7 | 7 |
| 4 | 5 | 5 | 64 | 61 | 15 | 7 | 10 |
| 5 | 4 | 5 | 37 | 40 | 5 | 5 | 7 |
| 6 | 5 | 5 | 43 | 40 | 5 | 5 | 10 |
| 7 | 6 | 5 | 79 | 81 | 15 | 7 | 8 |
| 8 | 5 | 5 | 48 | 48 | 15 | 7 | 8 |
| 9 | 5 | 5 | 99 | 98 | 15 | 7 | 9 |
| 10 | 4 | 5 | 26 | 30 | 5 | 5 | 5 |
| 11 | 7 | 5 | 37 | 35 | 25 | 9 | 7 |
| 12 | 4 | 5 | 24 | 22 | 15 | 7 | 9 |
| 13 | 8 | 5 | 22 | 22 | 45 | 13 | 9 |
| 14 | 1 | 5 | 15 | 17 | 5 | 5 | 7 |
| 15 | 5 | 5 | 70 | 71 | 5 | 5 | 6 |
| 16 | 6 | 5 | 45 | 38 | 15 | 7 | 13 |

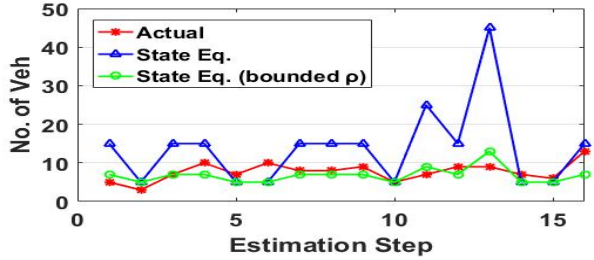

Fig. 1. Impact of Applying a Lower Bound in the State Equation Estimates.

*C. Proposed Estimation Approaches*

This section describes *two estimation approaches*. The first approach uses only probe vehicle data with a *fixed ρ* along the estimation intervals (e.g., $\rho$ = 20%), while the second approach utilizes fused data (probe vehicle and single loop detector data), where the single loop detector provides the estimation model with actual $\rho$ values.

*1) First Approach: Probe Data Assuming Fixed LMP (ρ)*

The $\rho$ value is an important variable in the proposed estimation method since it scales the probe measurements to reflect the total flow. The first estimation approach uses a fixed $\rho$ value in the estimation steps, observed from *historical* data. It should be noted that producing accurate estimates of the actual $\rho$ values will produce perfect estimation outcomes as long as there are no errors in the data; the $\rho$ values can be observed every time interval with the addition of a fixed sensor (e.g., a traditional loop detector or video detection system) to measure the total flow.

The predefined $\rho$ value in Equations (8) and (9) is computed as the arithmetic mean of all $\rho$ observations. For instance, Fig. 2 shows the actual $\rho$ values versus the predefined $\rho$ value as the red line ($\rho$ = 20%). This figure clearly shows errors in the in the $\rho$ value estimate. However, the KF reduces the errors produced from the fixed $\rho$ assumption, as shown later in the Results section. The varying $\rho$ values that appear in the figure is evidence that the proposed algorithm works well with noisy data.

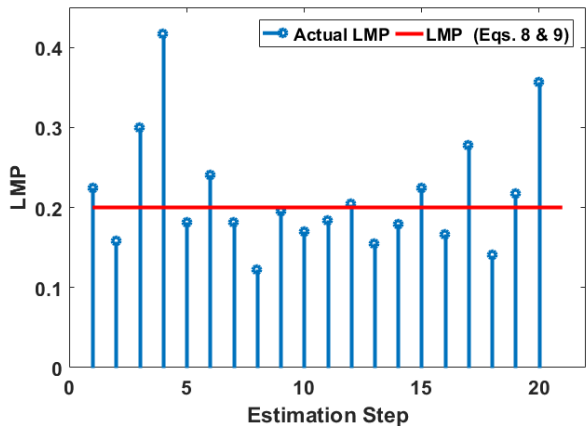

Fig. 2. Variation in actual LMP over all the estimation intervals.

*2) Second Approach: Fusion Data with Variable LMP (ρ)*

This section employs a single loop detector in addition to probe vehicle data to estimate the LMP. The loop detector is used to observe the actual total flow and then compute the $\rho$ values in the measurement equation for all the estimation times ($t$). The loop detector produces the total traffic flow, which can then be used to compute ($N_{total}$) in Equation (2). The new proposed estimation approach uses the same equations in the formulation section except that the measurement equation considers the actual $\rho$ values in the $H$ vector, as shown in Equation (16).

$$H(t) = \frac{2 \times \rho(t)}{q^{in}(t) + q^{out}(t)} \quad (16)$$

## IV. DATA COLLECTION

This paper evaluates the proposed estimation approaches using both empirical and simulated data. This section will describe both the data and the approach characteristics. The simulated data were used to provide additional data to test the algorithm for varying conditions (e.g., traffic demand levels and approach lengths).

*A. Empirical Data*

Fig. 3 shows the tested approach in downtown Blacksburg, VA. The approach falls between two traffic signals. The two observer locations define the approach length, as shown in Fig. 3. The approach length is approximately 74 meters based on Google Maps, and the speed limit is 25 mi/h (40 km/h). The two observers recorded the time stamp that each vehicle passed them. Using the data, it was possible to conduct a Monte Carlo simulation to extract a random sample of probe vehicles to compute ($q^{in}$ and $q^{out}$) and use them in Equation (15).

The team went to the field to collect actual field data for 75 minutes on March 29, 2018, between 4:00 and 5:15 p.m., observing a total of 813 vehicles. The full data served as the ground truth values, and thus our estimation method outputs were compared to these actual values. The collected data also included the observed travel time between Observer 1 and Observer 2 (from the beginning to the end of the approach). The tested approach had no roadway entrances or exits between the two observers and thus the flow continuity was maintained.

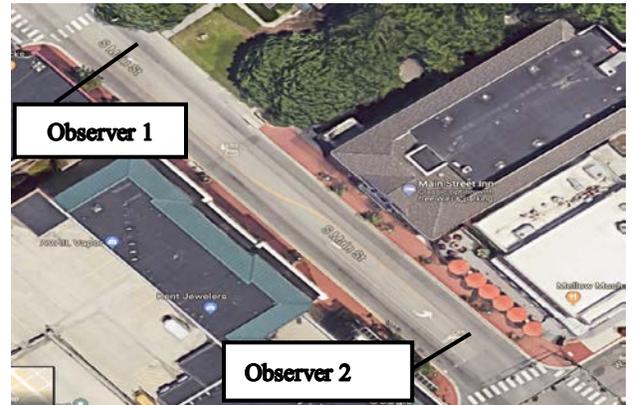

Fig. 3. Tested approach in Downtown Blacksburg, VA. (Source: Google Maps)

*B. Simulation Data*

INTEGRATION [26-29], a microscopic traffic simulation software, was used to validate and test the accuracy of the proposed method. The INTEGRATION software has been extensively validated and demonstrated to replicate empirical observations [30-35]. Specifically, INTEGRATION was used to create synthetic data for conditions not observed in the field to quantify the sensitivity of the proposed method to the approach length and traffic demand level. Specifically, a range of approach lengths were tested (i.e., 74, 150, 200, 300, and 400 m), as shown later in the Results section. The approach characteristics were calibrated to local conditions using typical values, which included a free-flow speed of 40 km/h, a speed-at-capacity 32 km/h, a jam density of 160 veh/km/lane, and a base saturation flow rate of 1800 veh/h/lane, which resulted in a roadway capacity of 855 veh/h given the cycle length and green times of the traffic signal. The traffic signal operated at a cycle length of 120 s and a 50:50 phase split. The amber and all-red interval was 3 s. These values were consistent with what was coded in the field.

## V. Results and discussion

The accuracy of the proposed KF estimator was tested using real and simulated data. The evaluation of all scenarios was based on the Relative Root Mean Square Error (RRMSE) and the Root Mean Square Error (RMSE), shown in Equations (17) and (18), respectively. The two measures are frequently used in the literature to compute the difference between the model estimates and the actual values.

$$RRMSE\ (\%) = \frac{100\sqrt{S\sum_{s=1}^{S}[\hat{N}^+(s) - N(s)]^2}}{\sum_{s=1}^{S}N(s)} \quad (17)$$

$$RMSE\ (veh) = \sqrt{\frac{\sum_{s=1}^{S}[\hat{N}^+(s) - N(s)]^2}{S}} \quad (18)$$

Where $N(s)$ represents the actual values, $\hat{N}^+(s)$ represents the estimated vehicle count values, and $S$ is the total number of estimations. The simulation starts with an erroneous initial estimation $\hat{N}^+(0) = 5$ veh while the real number is zero as in [9], an initial posterior estimate error $\hat{P}^+(0) = 5$ veh$^2$, and the measurement error covariance ($R$) is assumed to be 5 veh$^2$.

The proposed approach was evaluated using different probe vehicle LMPs, including 10, 20, 30, 40, 50, 60, 70, 80, and 90%.

### A. Empirical Data

#### 1) Sample Size Impact on Algorithm Performance

In this study, the estimation time interval was defined as the time when a prescribed number of probe vehicles traversed the approach (vehicles passed the traffic signal stop bar)—representing the desired sample size ($n$). This new approach ensures that the same number of probe vehicles are used every updating estimation time interval. First, an optimal sample size ($n$) is needed in the estimation equations. Different sample size values were tested (i.e., 1, 2, 3, 4, 5, 6, 7, 8, 9, and 10). In the proposed approach, the sample size ($n$) was used to identify the estimation time step; producing a variable estimation time step. TABLE II presents the RRMSE values for the tested sample sizes, RRMSE values for some sample sizes are close, especially in the values between five and eight. Consequently, the sample size ($n$) can be different depending on the data. In this section, the optimal sample size was defined to be *five*, once the fifth vehicle passed the second observer, the *TT* variable (the arithmetic mean travel time for the five vehicles) was updated in Equation (13).

TABLE II
RRMSE VALUES FOR 10 DIFFERENT SAMPLE SIZES FOR DIFFERENT LMPS

| Sample Size | LMP = 10% | LMP = 50% | LMP = 80% |
|---|---|---|---|
| 1 | 43 | 34 | 19 |
| 2 | 41 | 32 | 19 |
| 3 | 40 | 33 | 20 |
| 4 | 40 | 33 | 25 |
| 5 | 38 | 32 | 20 |
| 6 | 40 | 32 | 20 |
| 7 | 40 | 32 | 20 |
| 8 | 40 | 33 | 20 |
| 9 | 40 | 37 | 22 |
| 10 | 39 | 38 | 24 |

#### 2) Variable vs. Fixed Estimation Time Interval

Previous research always considers a constant estimation time step (e.g., 20 s). This is an appropriate approach if the entire data set is available (LMP of 100%) and/or the traffic demand is high. However, it is impossible to access the entire data set when dealing with probe vehicles. Consequently, a variable estimation time step is used rather than a fixed one.

This section demonstrates the benefits of using variable time steps as opposed to constant values, as is done in the literature. TABLE III presents the RRMSE values using variable and fixed estimation time steps. The proposed variable time step method was compared to the traditional fixed interval method (i.e., 15, 20, 30, 40, 50, 60, 120, and 240 s), with the results in TABLE III showing that low LMPs produce infinite values from the $H$ vector in Equation (8) in most fixed intervals. Consequently, the system produces NaN (Not a Number) values in Equation (17) due to lack of *TT* data.

TABLE III
RRMSE VALUES USING VARIABLE AND FIXED ESTIMATION INTERVAL TIME PERIODS

| Time Interval (s) | RRMSE (%) | | |
|---|---|---|---|
| | LMP = 20% | LMP = 50% | LMP = 80% |
| 15 | NaN | NaN | NaN |
| 20 | NaN | NaN | NaN |
| 30 | NaN | NaN | 91 |
| 40 | NaN | NaN | 96 |
| 50 | NaN | NaN | 43 |
| 60 | NaN | 57 | 40 |
| 120 | 63 | 49 | 36 |
| 240 | 59 | 60 | 39 |
| Proposed Algorithm (variable time interval) | 36 | 32 | 20 |

TABLE IV shows the average and maximum time interval for different LMPs. The results demonstrate that low LMPs require long intervals (e.g., 300 s) to ensure that some probe vehicles are on the approach. In contrast, approaches with high LMPs use short estimation intervals (e.g., 30 s). The proposed strategy ensures flexibility of the estimation intervals. Fig. 4 shows the relationship between the sample size (*n*) and the average time interval for different LMPs. It should be noted here that as the sample size (*n*) increases, the time interval increases. This approach ensures that a sufficient number of observations are available to estimate the traffic stream density within a desired margin of error. However, a smaller sample size can be used when faster computations are needed. The user can make the trade-off between the time interval and associated error they are willing to accept. In conclusion, the proposed algorithm enhances the estimation by reducing the estimation errors and allowing the algorithm to respond more quickly for high LMPs and ensuring that sufficient observations are available to compute the LMP for low LMPs.

TABLE IV
ESTIMATION TIME INTERVAL FOR DIFFERENT LMPs

| LMPs % | Avg Time Interval (s) | Max Time Interval (s) |
|---|---|---|
| 10 | 254 | 450 |
| 20 | 131 | 234 |
| 30 | 86 | 177 |
| 40 | 66 | 123 |
| 50 | 53 | 111 |
| 60 | 43 | 99 |
| 70 | 37 | 101 |
| 80 | 33 | 97 |
| 90 | 29 | 78 |

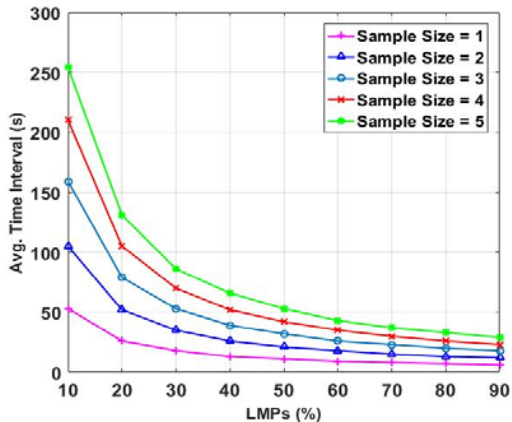

Fig. 4. Impact of sample size on the average time interval.

*3) Impact of Probe Vehicle LMPs*

The number of vehicles along the approach was estimated considering different LMPs (i.e., 10, 20, 30, 40, 50, 60, 70, 80, and 90%). A Monte Carlo simulation was conducted to create 100 random samples from the collected data for each scenario, creating a random sample of probe vehicles. The proposed algorithm used the optimal sample size obtained in the previous section (*n*=5 veh). The more vehicle information available (i.e., the higher the LMP), the shorter the estimation interval and the more estimation steps are possible. Fig. 5 presents the estimation along different LMPs for the empirical data. The RRMSE values produced reasonable values even with low LMPs, as shown in TABLE V. For instance, the estimated vehicle count values are off by 2.8 vehicles when the LMP is equal to 10%. On the other hand, the estimated vehicle count values are off by 1.0 vehicle when the LMP is equal to 90%. From TABLE V, it is clear that the error increases as the LMP decreases. It should be noted that the total field-collected data was not enough to test the model accuracy at low LMPs. Thus, the model's accuracy was further tested using simulated data.

TABLE V
RRMSE AND RMSE VALUES FOR NINE SCENARIOS USING VARIOUS LMPs FOR REAL DATA

| LMPs % | RRMSE (%) | RMSE (veh) |
|---|---|---|
| 10 | 38 | 2.8 |
| 20 | 36 | 2.6 |
| 30 | 35 | 2.5 |
| 40 | 34 | 2.4 |
| 50 | 32 | 2.3 |
| 60 | 28 | 2.0 |
| 70 | 25 | 1.8 |
| 80 | 20 | 1.4 |
| 90 | 14 | 1.0 |

*4) Impact of Fusing Probe and Single Loop Detector Data*

This section considers computes the actual $\rho$ in the measurement equation (Equation (16)) using a stationary sensor in addition to the probe data. The $\rho$ value is defined as the ratio of the number of probe vehicles to the total number of vehicles, as in Equation (2). The RRMSE values using the developed two estimation approaches, namely: (1) Probe vehicle data approach assuming fixed $\rho$ values (named probe approach), and (2) probe vehicle and single loop detector data using variable $\rho$ values (named fusion approach) are shown in TABLE VI. Different loop detector locations were tested for the fusion approach (entrance, exit, and middle) to measure the actual $\rho$ values Based on the RRMSE values, in some cases, installing a loop detector in the middle of the tested approach would slightly improve the model's accuracy at low LMPs by up to 4%; however, installing a loop detector may not be cost effective. In conclusion, we recommend using data from existing detection sensors (e.g., loop detectors or video surveillance) if they already exist on the roads, otherwise, we recommend using a fixed $\rho$ value that can be estimated from historic data rather than the actual $\rho$.

TABLE VI
RRMSE VALUES USING ONE-LOOP DETECTOR IN DIFFERENT LOCATIONS (ENTRANCE, MIDDLE, AND EXIT)

| LMPs % | RRMSE (%) | | | |
|---|---|---|---|---|
| | Probe Approach (Fixed $\rho$) | Fusion Approach (Variable $\rho$) | | |
| | | Entrance | Middle | Exit |
| 10 | 38 | 37 | 34 | 38 |
| 20 | 36 | 35 | 33 | 36 |
| 30 | 35 | 34 | 32 | 36 |
| 40 | 34 | 34 | 33 | 34 |
| 50 | 32 | 31 | 31 | 32 |
| 60 | 28 | 28 | 28 | 28 |
| 70 | 25 | 25 | 24 | 25 |
| 80 | 20 | 19 | 19 | 20 |
| 90 | 14 | 14 | 12 | 14 |

*B. Simulation Data*

The simulation software used the same approach characteristics that were observed during the data collection in downtown Blacksburg. This section first investigates the sensitivity of the vehicle count estimation model to the approach length and to the traffic demand level. Then, the simulated data was used to examine the effect of the choice of the $\rho$ value on the estimation accuracy. First, the estimation model is tested considering a constant predefined $\rho$ in the model equations (probe approach). Second, the model accuracy is examined using the actual $\rho$ obtained from the installation of a single loop detector (fusion approach). In this section, the optimal sample size for simulated data was *eight*.

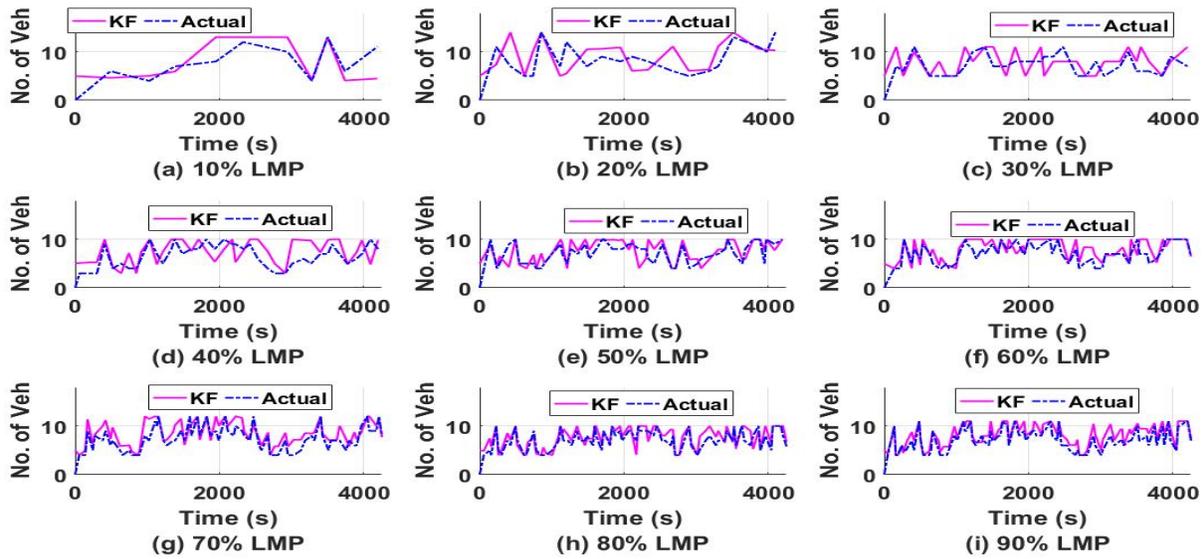

Fig. 5. Actual and estimated vehicle counts over estimation intervals.

*1) Approach Length and Traffic Demand Sensitivity Analysis*

First, the simulated data were used to study the sensitivity of the estimation model to the approach length. Different approach lengths were investigated in addition to the original approach length (i.e., 150, 200, 300, and 400 m). TABLE VII presents the RRMSE values for different approach lengths for different LMPs. The results demonstrate that the estimation accuracy increases with an increase in the approach length, which is in line with Vigos *et al.*'s conclusion [9]. For the rest of this study, we used a 400m approach length to ensure that the approach accommodates more vehicles.

TABLE VII
RRMSE VALUES UNDER DIFFERENT APPROACH LENGTHS

| Approach Length | LMP = 20% | LMP = 50% | LMP = 80% |
| --- | --- | --- | --- |
| 74 | 39 | 37 | 29 |
| 150 | 36 | 30 | 19 |
| 200 | 33 | 26 | 16 |
| 300 | 29 | 22 | 13 |
| 400 | 25 | 21 | 13 |

Second, the impacts of traffic demand level on the estimation model were then examined, considering both under- and over-saturated conditions. Different v/c (flow/capacity) ratios were tested (from 0.1 to 1.1 at 0.1 increments) as shown in TABLE VIII. The original v/c ratio was 0.79 (650/855 = 0.79) based on the collected data. In general, the RMSE and the RRMSE decrease with the increase of LMPs for the same traffic demand level. However, the RMSE is expected to increase with increasing traffic demand levels for the same LMP, the reason behind that is the increment in the total number of vehicles on the tested approach. For instance, at 10% LMP, for the 0.2 v/c, the RRMSE is 59%, and the RMSE is 2.0, so we are off by 2.0 out of the actual 3.4 vehicles, while for a 1.0 v/c ratio, the RRMSE is 24%, and the RMSE is 6.5, so we are off by 7.0 vehicles out of the actual 26 vehicles. In conclusion, the RMSE value can be higher but represents better results (in our case the total number of vehicles). The results from the table demonstrate that the estimation model works better as the level of congestion increases (e.g., v/c of 0.9, 1.0, and 1.1). The proposed model therefore demonstrates the model's efficiency with over-saturation scenarios (e.g., v/c = 1.1), especially for low LMPs, indicating its usefulness within a real-time traffic signal controller. Accordingly, 1.1 v/c ratio is used in the next section given that real-time traffic signal control is mostly needed during congested periods.

In the next results sections, the simulated data were employed to examine the effect of the choice of $\rho$ on the estimation accuracy, namely: using a constant $\rho$ versus using the actual $\rho$ that could be obtained if a single loop detector was installed.

TABLE VIII
RRMSE AND RMSE VALUES FOR DIFFERENT V/C RATIOS

| LMPs % | RRMSE (%), RMSE (veh) | | | | | | | | | | |
|---|---|---|---|---|---|---|---|---|---|---|---|
| | V/C = 0.1 | V/C = 0.2 | V/C = 0.3 | V/C = 0.4 | V/C = 0.5 | V/C = 0.6 | V/C = 0.7 | V/C = 0.8 | V/C = 0.9 | V/C = 1.0 | V/C = 1.1 |
| 10 | 75, 0.8 | 59, 2.0 | 48, 2.7 | 43, 3.2 | 36, 3.4 | 33, 3.9 | 34, 4.7 | 29, 5.3 | 25, 6.3 | 24, 6.5 | 16, 5.1 |
| 20 | 73, 1.2 | 58, 1.9 | 45, 2.5 | 40, 3.1 | 34, 3.2 | 32, 3.7 | 29, 4.3 | 27, 5.0 | 23, 5.9 | 23, 6.3 | 14, 4.7 |
| 30 | 73, 1.2 | 55, 1.8 | 44, 2.4 | 40, 3.0 | 33, 3.1 | 29, 3.5 | 27, 4.0 | 26, 4.7 | 23, 5.7 | 22, 5.8 | 13, 4.4 |
| 40 | 73, 1.2 | 55, 1.8 | 41, 2.2 | 36, 2.7 | 30, 2.9 | 28, 3.3 | 26, 3.8 | 24, 4.3 | 21, 5.3 | 20, 5.4 | 13, 4.4 |
| 50 | 69, 1.2 | 46, 1.5 | 40, 2.2 | 33, 2.5 | 28, 2.7 | 26, 3.1 | 24, 3.5 | 22, 4.0 | 19, 4.8 | 18, 4.7 | 13, 4.4 |
| 60 | 59, 1.0 | 43, 1.4 | 35, 1.9 | 30, 2.3 | 25, 2.4 | 23, 2.7 | 20, 3.0 | 18, 3.4 | 15, 3.9 | 15, 4.1 | 12, 3.9 |
| 70 | 52, 0.9 | 42, 1.3 | 32, 1.8 | 27, 2.0 | 22, 2.1 | 20, 2.3 | 16, 2.4 | 15, 2.8 | 13, 3.3 | 13, 3.5 | 10, 3.4 |
| 80 | 48, 0.8 | 35, 1.2 | 29, 1.6 | 24, 1.8 | 19, 1.8 | 16, 1.9 | 13, 2.0 | 14, 2.5 | 11, 2.8 | 11, 3.0 | 9, 2.9 |
| 90 | 45, 0.8 | 28, 1.0 | 24, 1.3 | 22, 1.6 | 16, 1.5 | 14, 1.6 | 11, 1.5 | 11, 2.0 | 9, 2.2 | 9, 2.4 | 9, 2.9 |

*2) Probe Vehicle Impact on Algorithm Performance using Fixed ρ Values*

The proposed estimation model was evaluated using simulation data. Again a Monte Carlo simulation was run to create 100 samples from the full data set for each scenario. In this approach, we assume the ratio between the number of probe vehicles and the number of total vehicles is constant. The estimation equations use a predefined fixed $\rho$ value (e.g., an average value from historical data). TABLE IX presents the RRMSE and RMSE values using the simulation data. The RRMSE values produced reasonable values even with low LMPs, as shown in TABLE IX. For instance, the vehicle count estimates were off by 16% when the LMP equaled 10%. On the other hand, our vehicle count estimates values were off by 9% for LMPs of 90%. Furthermore, the vehicle count model produced RMSE values of up to 5.1 vehicles. Knowing that the tested approach can accommodate up to 64 vehicles based on the jam density value, these RMSE values are low. Fig. 6 presents the estimation at different LMPs. As a result, the proposed method addresses the research goal appropriately, producing reasonable error values. Vigos *et al.* considered their model a robust model with up to a 27.5% RRMSE using at least three loop detector measurements [9]. It is obvious that using the predefined $\rho$ values results in errors. However, the KF is able to reduce these errors.

TABLE IX
RRMSE AND RMSE VALUES FOR VARIOUS LMPs

| LMPs % | RRMSE (%) | RMSE (veh) |
|---|---|---|
| 10 | 16 | 5.1 |
| 20 | 14 | 4.7 |
| 30 | 13 | 4.4 |
| 40 | 13 | 4.4 |
| 50 | 13 | 4.4 |
| 60 | 12 | 3.9 |
| 70 | 10 | 3.4 |
| 80 | 9 | 2.9 |
| 90 | 9 | 2.9 |

*3) Fusion Data Impact on Algorithm Performance using Variable ρ Values*

This section compares the two estimation approaches: The probe approach and the fusion approach. Again, different loop detector locations were tested for the fusion approach (entrance, exit, and middle) to measure the actual $\rho$ values. The variation of the RRMSE considering a stationary sensor (e.g., loop detector) are shown in TABLE X for different traffic demand levels (*v/c* ratio of 0.2, 0.5, and 1.1). Based on the RRMSE values, in some scenarios, installing a loop detector in the middle of the tested approach would slightly improve the model's accuracy by up to 8%; however, installing a loop detector may not be cost effective. In conclusion, we recommend using a fixed $\rho$ value that can be estimated from historic data rather than the actual $\rho$. Summary and Conclusions

This research proposed a novel algorithm for estimating the number of vehicles on a signalized approach using probe vehicle data only. The proposed estimation model uses a variable estimation interval that ensures a predefined number of probe vehicles are observed in each estimation interval. The estimation equations use the linear KF technique. The state-space equation is based on the conservation equation while the travel time measurements together with the hydrodynamic equation are used to construct the measurement equation. Two estimation approaches were presented, namely: (a) using only the probe vehicles' data to estimate the vehicle counts; this approach uses a predefined LMP value (obtained from historical data) (b) using a single loop detector located somewhere near the middle of the section to estimate the actual LMP values. The model estimation accuracy was evaluated using both empirical data (collected in downtown Blacksburg, VA) and simulated data.

The work done for this paper demonstrates the importance of having a variable estimation interval, and its benefits on the estimation accuracy, especially when dealing with low LMPs. In computing the estimation interval, the algorithm first needs a certain sample size (*n*) to be defined.

The study also investigated the sensitivity of the model to the approach length and traffic demand level, showing that the model's relative accuracy increases as the approach length increases given that the number of vehicles increases. The study also examined different demand levels (*v/c* ratios) in order to evaluate the model's efficiency, with results showing that dealing with a high traffic demand levels improved the estimation model.

In both estimation approaches, the results show that the estimation error increases as the LMP decreases. In some scenarios, the second approach (real-time estimated LMP) produces smaller errors since the actual LMP values can be observed. However, use of the second approach is not recommended, as it adds only slight improvements to the estimation outcomes with the additional cost associated with installing a loop detector, which may be cost prohibitive especially in large urban areas.

Proposed future work entails developing on-line learning techniques to enhance the estimation of the $\rho$ values.

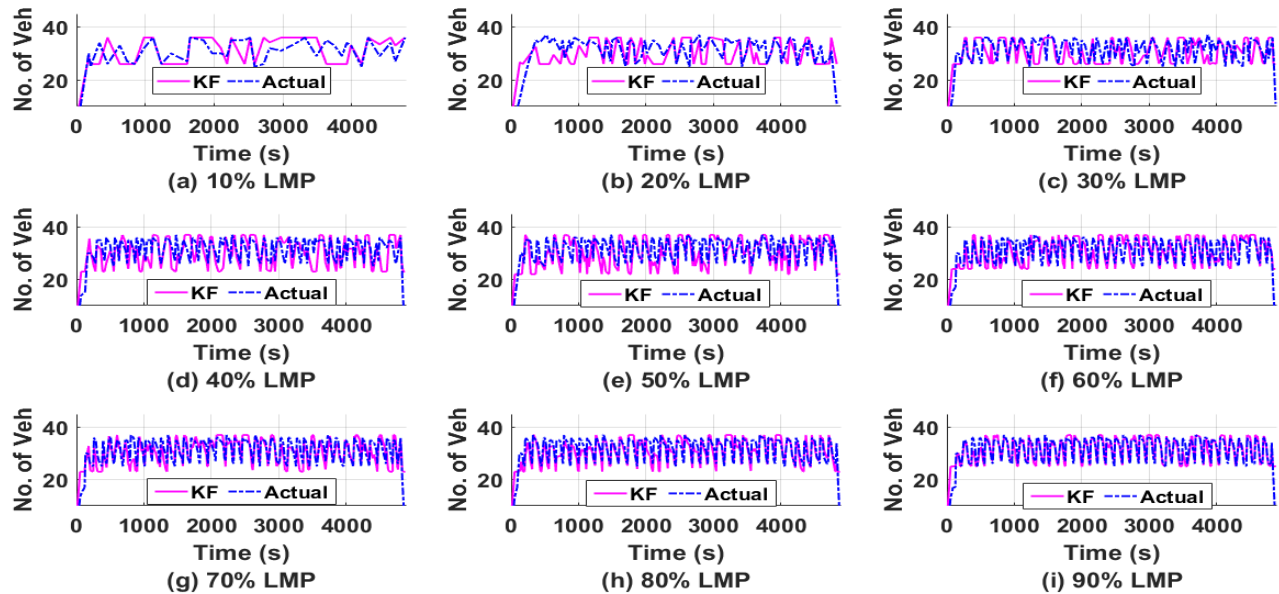

Fig. 6. Actual and estimated vehicle counts using simulated data.

TABLE X
RRMSE VALUES USING ONE-LOOP DETECTOR IN DIFFERENT LOCATIONS (ENTRANCE, MIDDLE, AND EXIT)

| LMPs % | V/C = 0.2 | | | | V/C = 0.5 | | | | V/C = 1.1 | | | |
|---|---|---|---|---|---|---|---|---|---|---|---|---|
| | Probe Approach | Fusion Approach | | | Probe Approach | Fusion Approach | | | Probe Approach | Fusion Approach | | |
| | | Entrance | Middle | Exit | | Entrance | Middle | Exit | | Entrance | Middle | Exit |
| 10 | 59 | 55 | 51 | 53 | 36 | 35 | 34 | 37 | 16 | 15 | 14 | 15 |
| 20 | 58 | 54 | 51 | 53 | 34 | 33 | 33 | 34 | 14 | 15 | 15 | 15 |
| 30 | 55 | 53 | 51 | 53 | 33 | 32 | 32 | 33 | 13 | 14 | 14 | 14 |
| 40 | 55 | 49 | 50 | 51 | 30 | 30 | 30 | 30 | 13 | 15 | 15 | 16 |
| 50 | 46 | 47 | 46 | 48 | 28 | 27 | 27 | 28 | 13 | 14 | 13 | 14 |
| 60 | 43 | 43 | 43 | 43 | 25 | 25 | 25 | 25 | 12 | 12 | 12 | 12 |
| 70 | 42 | 39 | 39 | 40 | 22 | 21 | 21 | 22 | 10 | 10 | 10 | 11 |
| 80 | 35 | 34 | 34 | 36 | 19 | 19 | 20 | 19 | 9 | 8 | 8 | 9 |
| 90 | 28 | 27 | 27 | 28 | 16 | 16 | 16 | 16 | 9 | 9 | 9 | 9 |

## VI. ACKNOWLEDGMENTS

This effort was funded by the Urban Mobility and Equity Center (UMEC).

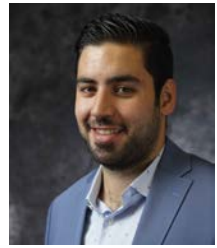

**Mohammad A. Aljamal** (M'18) received his B.Sc. degree in civil engineering from Jordan University of Science and Technology, Irbid, Jordan, in 2014 and his M.Sc. degree in civil engineering from Virginia Tech University, Blacksburg, VA, USA, in 2017. He is currently a Ph.D. candidate in the Department of civil engineering at Virginia Tech. He works at the Center for Sustainable Mobility at the Virginia Tech Transportation Institute. His research interests include traffic flow theory, intelligent transportation systems, machine learning, evacuation modeling, traffic modeling and simulation. Mohammad is a member of IEEE, member of ITE, member of Council of International Transportation Engineers at Virginia Tech, and a member of Jordanian Engineers Association (JEA).

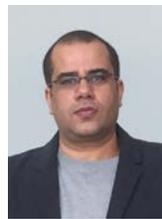

**Hossam M. Abdelghaffar** received his B.Sc. degree in Electronics Engineering from Faculty of Engineering, Mansoura University, Egypt, M.Sc. degree in Automatic Control System Engineering, Mansoura University, Egypt, and Ph.D. degree from the Bradley Department of Electrical and Computer Engineering at Virginia Tech. He is currently an assistant professor in the Department of Computers & Control Systems, Faculty of Engineering, Mansoura University, Egypt. In addition, he is a Postdoctoral Associate with the Center for Sustainable Mobility at the Virginia Tech Transportation Institute, Virginia Tech.

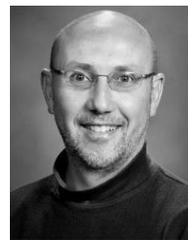

**Hesham Rakha** (M'04, SM'18, F'20) received his B.Sc. (with honors) in civil engineering from Cairo University, Cairo, Egypt, in 1987 and his M.Sc. and Ph.D. in civil and environmental engineering from Queen's University, Kingston, ON, Canada, in 1990 and 1993, respectively. Dr. Rakha's research focuses on large-scale transportation system optimization, modeling and assessment. Specifically, Dr. Rakha and his team have expanded the domain of knowledge (in traveler and driver behavior modeling) and developed a suite of multi-modal agent-based transportation modeling tools, including the INTEGRATION microscopic traffic simulation software. This software was used to evaluate the first dynamic route guidance system, TravTek in Orlando, Florida; to model the Greater Salt Lake City area in preparation for the 2002 Winter Olympic Games; to model sections of Beijing in preparation for the 2008 Summer Olympic Games; to optimize and evaluate the performance of alternative traveler incentive strategies to reduce network-wide energy consumption in the Greater Los Angeles area; and to develop and test an Eco-Cooperative Automated Control (Eco-CAC) system. Finally, Dr. Rakha and his team have developed various vehicle energy and fuel consumption models that are used world-wide to assess the energy and environmental impacts of ITS applications, including the VT-Micro, VT-Meso, the VT-CPFM, the VT-CPEM, and the VT-CPHEM models.